\documentclass[prl,twocolumn,superscriptaddress,showpacs,nobalancelastpage,amsmath,amssymb]{revtex4}
\usepackage{graphicx}% Include figure files
\usepackage{bm}% bold math
\begin{document}
\date{\today}
%opening
\title{
One-dimensional weak antilocalization and band Berry phases in HgTe wires
}

\author{M. M\"uhlbauer, A. Budewitz, B.\ B\"uttner, G. Tkachov, E.M.~Hankiewicz, C. Br\"une, H.\ Buhmann, and L.W.\ Molenkamp}

\affiliation{
Faculty of Physics and Astrophysics, University of W\"urzburg, Am Hubland, 97074 W\"urzburg, Germany}

\begin{abstract}
We study the weak antilocalization (WAL) effect in the magnetoresistance of narrow HgTe wires fabricated in quantum wells (QWs) with normal and inverted band ordering. Measurements at different gate voltages indicate that the WAL is only weakly affected by Rashba spin-orbit splitting and persists when the Rashba splitting is about zero. The WAL signal in wires with normal band ordering is an order of magnitude smaller than for inverted ones. These observations are attributed to a Dirac-like topology of the energy bands in HgTe QWs. From the magnetic-field and temperature dependencies we extract the dephasing lengths and band Berry phases. The weaker WAL for samples with a normal band structure can be explained by a non-universal Berry phase which always exceeds $\pi$, the characteristic value for gapless Dirac fermions.
\end{abstract}

\maketitle

%\section{Introduction}
{\em Introduction} -- Weak anti-localization (WAL) is a quantum transport effect that occurs in electronic systems with broken spin rotation symmetry. It is associated with the spin precession along closed electron trajectories, which leads to reduced backscattering and, in materials with sufficiently strong spin-orbit interaction (SOI), to an observable positive magnetoresistance \cite{Bergmann82}.
Whereas, in disordered systems without SOI, the related weak localization (WL) is a precursor for an insulating state, WAL, in marked contrast, indicates that the system remains metallic in the zero-temperature limit. While in semiconductors with a parabolic band dispersion, WAL results from the Rashba or Dresselhaus SOI (see e.g. \cite{Bergmann82,Dresselhaus92}), Suzuura and Ando suggested that the WAL may also occur in materials with a Dirac-like band dispersion, such as a two-dimensional (2D) honeycomb lattice \cite{Ando02}. In such systems the carrier backscattering is supposed to be absent due to a topological Berry phase of the carriers at the Fermi energy, which protects the carriers from localization.
However, in graphene, the first Dirac material to be extensively studied experimentally \cite{GeimKim05}, the band-structure induced WAL effect is obscured by a competing WL effect (see e.g. Refs.~\cite{Geim06,Tikhonenko09}) due to intervalley scattering \cite{McCann06}. In narrow graphene wires this scattering mechanism is ubiquitous due to the edge roughness, which can explain the absence of the WAL in recent transport measurements \cite{Minke12} and numerical simulations \cite{Ortmann11} for graphene nanoribbons. As topological effects in transport physics are in the focus of current research, it is of great interest to examine the quasi-1D localization properties of other novel materials exhibiting a Dirac-like band structure.

In this work we study quantum transport in nano\-wires fabricated from HgTe quantum wells (QWs).
HgTe QWs with thicknesses $d$ larger than a critical value $d_c$ exhibit an inverted band structure,
which gives rise to the formation of a quantum spin Hall insulator state, characterized
by gapless Dirac-like helical edge states when the Fermi energy is in the bulk gap \cite{Bernevig06,Koenig07}. Here, we are interested in the transport properties of the bulk conduction band states.
HgTe QWs with a thickness close to $d_c$ have a very small gap and the conduction band structure exhibits a Dirac-like dispersion, as found in band-structure calculations and verified by transport experiments \cite{Buettner11,Tkachov11a} and magneto-optical \cite{Zholudev12,Olbrich13} observations.
In contrast to graphene, HgTe QWs have only a single Dirac-like valley, so that intervalley scattering is absent, with obvious advantages for studying quantum interference effects of Dirac fermions. Therefore, in narrow HgTe QW wires a clear WAL effect is observable.

%%%  Figure 1   %%%
\begin{figure}[t!]
\includegraphics[width=85mm]{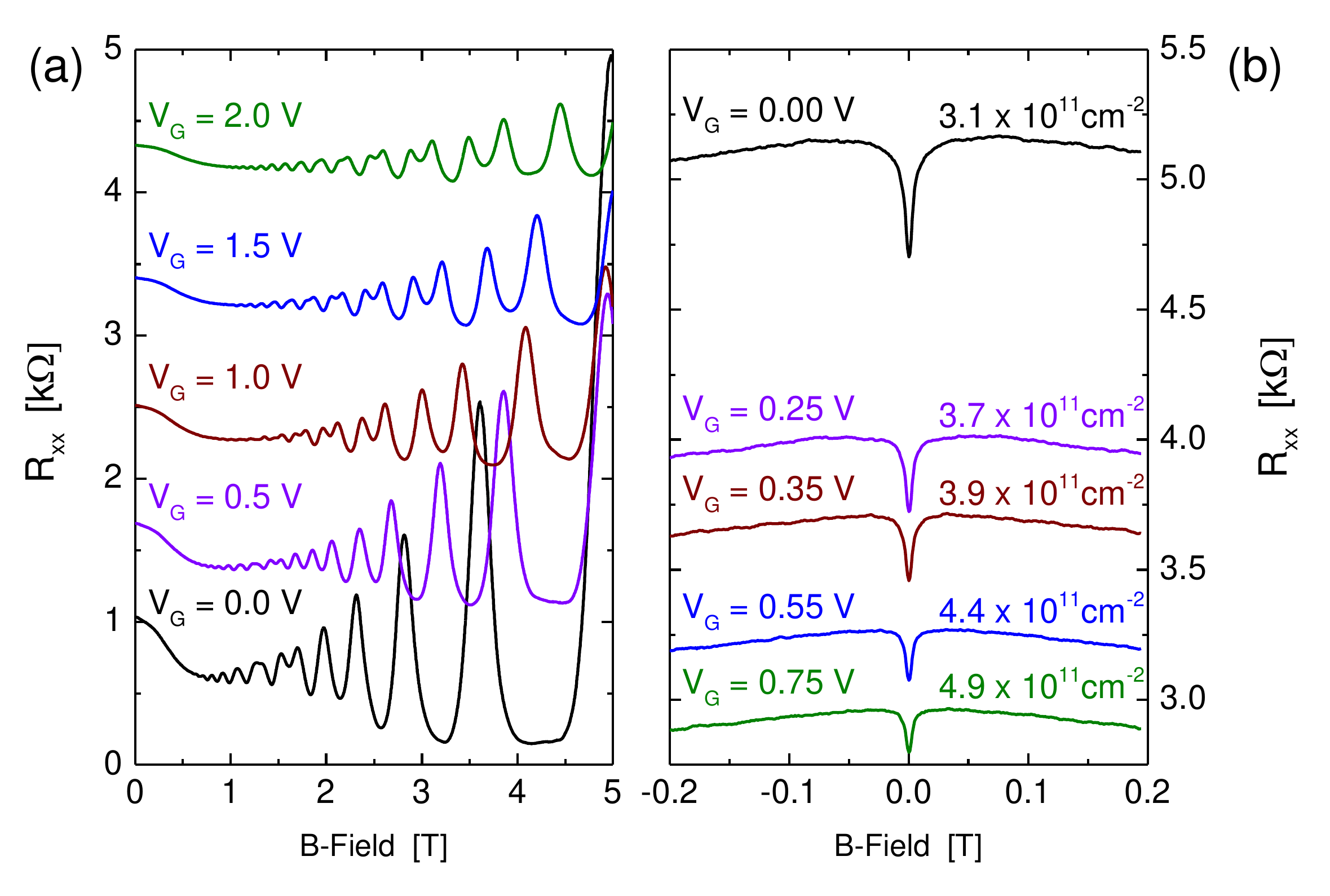}
\caption{	
$R_{xx}(B)$, of a macroscopic Hall bar (for clarity, traces at increased density are off-set by 1 k$\Omega$ with respect to the previous.) (a) and WAL resistance dip of the wire array sample for the 7 nm QW at different gate-voltages (b). The gate-voltages in (a) correspond to $n = 3.2$, 4.3, 5.4, 6.7, and $7.9 \times 10^{11}$ cm$^{-2}$ for 0.0 to 2.0 V, respectively.}
\label{R}
\end{figure}

%%%%%%%%%%%%%%%%%%%%%%%%%% Experiment %%%%%%%%%%%%%%%%%%%%%%%%%%%%%%%%%%%%%%%
%
%%%   Figure 2   %%%
%
%\begin{widetext}
\begin{figure*}%[b]
\includegraphics[width=180mm]{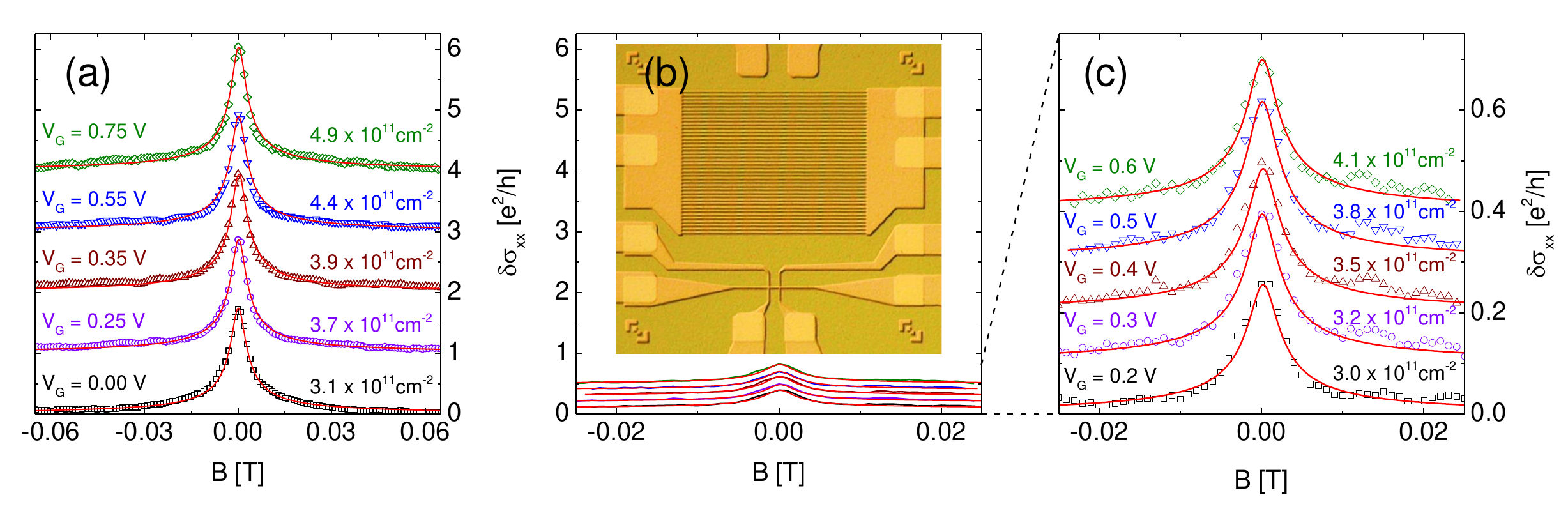}
\caption{
WAL contribution to the magnetoconductivity, $\delta\sigma_{xx}(B)$, for wires of the 7 nm (inverted) QW, Fig. (a), and of the 5 nm (normal) QW, Figs. (b) and (c) at different gate voltages $V_g$.
The thin (red) lines are fits based on Eqs. (\ref{dS}) - (\ref{l_M}).
Subsequent curves are shifted by $e^2/h$ in (a) and by $0.1 e^2/h$ in (b) and (c) for clarity.
The inset of Fig. (b) shows the mesa structure of measured samples.
}
\label{dS_B}
\end{figure*}
%\end{widetext}
%

{\em Experiment} --
We present experimental results on samples fabricated on two different QW structures. The structures were grown by molecular beam expitaxy on insulating CdTe and CdZnTe substrates with nominal QW thicknesses of 5 nm ($< d_c$) and 7 nm ($> d_c$), respectively, thus exhibiting a normal and an inverted band structure ordering. The HgTe QWs are embedded between thick Hg$_{0.3}$Cd$_{0.7}$Te barriers. The 7 nm QW sample is additionally doped with iodine in the bottom barrier at a distance of 70 nm. The layers have been patterned by optical and electron beam lithographical techniques into Hall bars with $30 \times 10$ $\mu$m$^2$ (for characterization purposes) and into an array of 40 parallel wires for the quantum interference investigations. The wires are 50~$\mu$m long and approximately 0.37~$\mu$m wide. All device structures are fitted with a Ti/Au gate electrode on a 50 nm thick SiO$_2$/Si$_3$N$_4$ multilayer gate insulator to control the carrier density and the structural inversion asymmetry and thus the Rashba SOI strength. Densities and mobility for the wires have been extrapolated from the Hall bar structures. A micrograph of the wire sample is given as inset in Fig.~2. By measuring a large number of identical wires in parallel, universal conductance fluctuations are averaged out while non-random quantum-interference effects like the WAL remain. All measurements have been carried out in a $^4$He cryostat fitted with a superconducting magnet at a base temperature of 1.8 K.

Figure~1 shows magnetotransport data for the 7 nm QW on a macroscopic Hall bar (a) and the wire sample (b) for five different gate voltages. The first observation is that the wire sample exhibits a sharp resistance dip around $B = 0$ riding on a smoothly varying parabolic background which we ascribe to the electron-electron correction to the conductivity. The peak height appears to be approximately constant in the presented gate-voltage range (Fig.~1b). The latter is remarkable since from the oscillations of the magnetoresistance (Fig.~1a) one can infer that in this gate-voltage range the Rashba SOI strength approaches zero for $V_g$ between 0.5 and 1.0~V - the beating pattern vanishes when the applied gate-voltage compensates the doping induced structural inversion asymmetry. A similar behavior is observed for the 5 nm QW.

In order to compare the WAL effect on the normal (5 nm) and inverted (7 nm) QW, the WAL peak was measured in both samples for gate voltages where the samples exhibit similar carrier densities. The interference-induced WAL signal, $\delta\sigma_{xx}$, is extracted from the measured magnetoconductance by subtracting the smooth parabolic background  and considering the device geometry. The results are presented in Fig.~2. The magnitude of the WAL peak is found between 1.8 and 2 $e^2/h$ for the inverted QW (Fig.~2a) and, remarkably, one order of magnitude smaller for the normal QW (Fig.~2b). For a better comparison we enlarged the conductivity scale for the normal QW (Fig.~2c ).  In both cases, the WAL features display a very similar line shape.
For further analysis we require the relevant phase coherence length for the samples under investigation. A convenient way to obtain this quantity is to measure the temperature dependence of the WAL peak. The results are plotted in Fig.~3a for five different temperatures, up to 16 K,  and three different gate voltages. As expected for a quantum interference effect the peak amplitude decreases exponentially with temperature for both QWs.

%%%%%%%%%%%%%%%%%%%%%%%%%%%%%%%%%   Model   %%%%%%%%%%%%%%%%%%%%%%%%%%%%%%%%%%%

\begin{figure}[t]
\includegraphics[width=80mm]{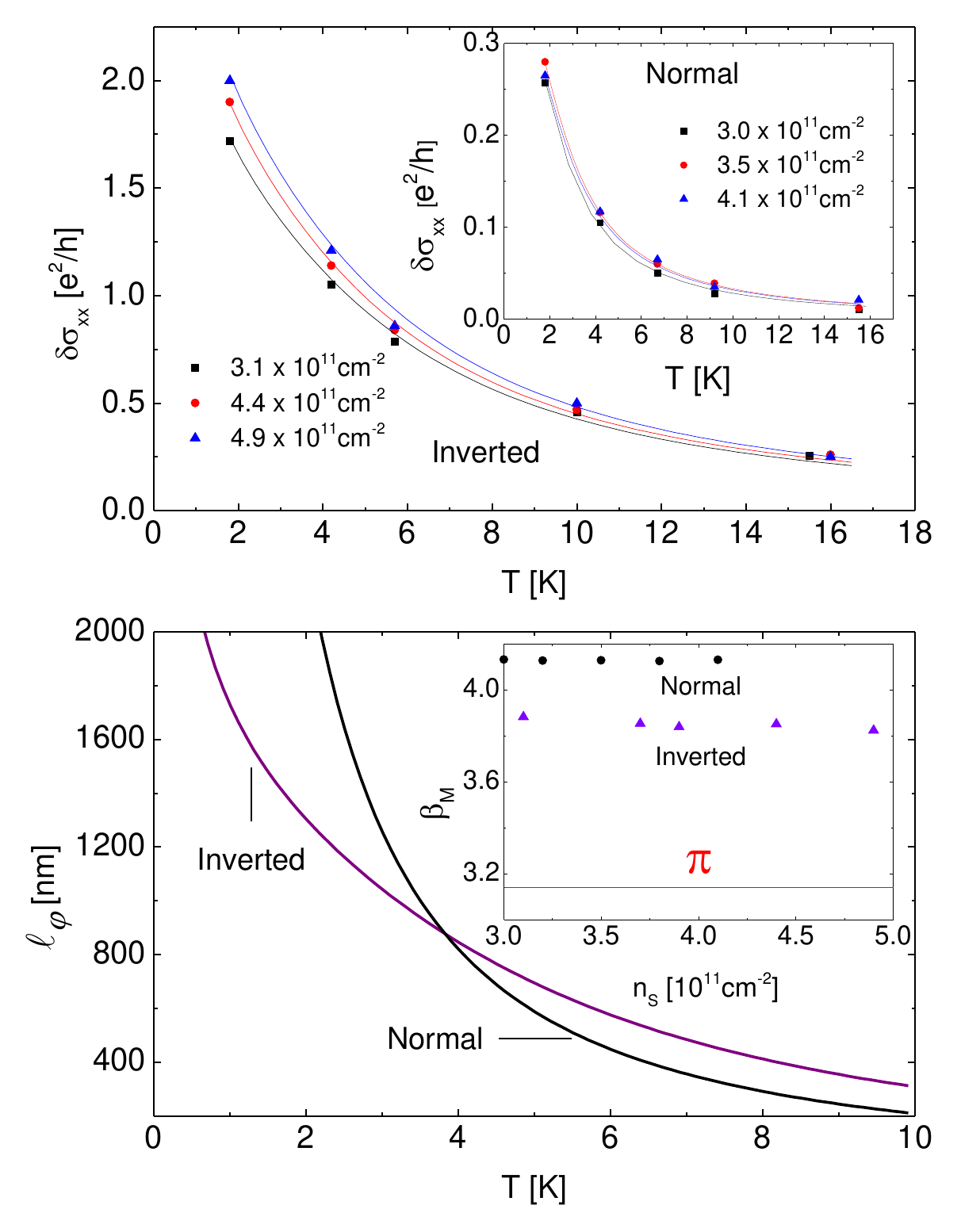}
\caption{
(a) Temperature dependence of the conductivity $\delta\sigma_{xx}(T)$ for wires of the 7 nm (inverted) QW and 5 nm (normal) QW (inset) at different gate voltages. 
The solid (red) lines are fits based on Eqs. (\ref{dS}) - (\ref{l_M}).
(b) Extracted dephasing length for wires of the inverted and normal QW at $V_g=0$.
Inset: Extracted Berry phase, $\beta_M$, versus carrier density $n_s$ for inverted and normal samples.
}
\label{dS_T}
\end{figure}

{\em Model} --
Since the magnitude of the WAL presented here is several times larger than in the corresponding 2D systems \cite{Olshan10,Minkov12}, we attribute the observed pronounced WAL to the quasi-1D diffusive character of the transport, i.e. the width of the sample, $w$, is significantly smaller than the phase coherence length, $\ell_\varphi$. We verify this conjecture by comparing the measurements with the theoretical results for the quantum-interference correction to the classical (Drude) conductivity, $\delta\sigma_{xx}$, which is different for different dimensions. The appropriate band structure parameter, $\cal{M}$, $\cal{B}$ and $\cal{D}$, we infer from calculation of the four-band Bernevig-Hughes-Zhang (BHZ) model for the HgTe QWs \cite{Bernevig06} and the approach of Ref. \cite{Tkachov11b} appropriately adapted to the wire geometry, $w \ll \ell_\varphi$. The conductivity correction is then given by
%
%%%   Equation 1   %%%
%
\begin{eqnarray}
\delta\sigma_{xx}=  \frac{ 2\frac{e^2}{h} \,\, (1 + C_M) }{ w\sqrt{ \ell^{-2}_\varphi + \ell^{-2}_B + \ell^{-2}_M   } },
\label{dS}
\end{eqnarray}
where the factor of 2 accounts for independent contributions of the Kramers-partner blocks in the BHZ Hamiltonian, $\ell_\varphi$, $\ell_B$ and $\ell_M$ are the characteristic length-scales for transport in the diffusive quantum interference regime:
\begin{eqnarray}
\ell_\varphi &=& (c_1 T^{2/3} + c_2 T^3)^{-1/2},\,\,
\ell_B=\sqrt{3} \hbar/(w |eB|),
\label{l_phi}
\\
\ell_M &=& \frac{ \ell }{\sqrt{2} |1 - \beta_M/\pi|}, \,\,
C_M \approx -\frac{19}{2} \left( 1 - \frac{\beta_M}{ \pi}  \right)^2,\,\,
\label{l_M}
\\
\beta_M &=&\pi \left(1 +  \frac{ {\cal M} + {\cal B} k^2_{_F} }{E_F - {\cal D} k^2_{_F}}  \right).
\label{b_M}
\end{eqnarray}
In Eq.~(\ref{l_phi}) $\ell_\varphi$ represents the dephasing length due to electron-electron and electron-phonon interactions with respective contribution $\propto T^{2/3}$ and $\propto T^3$ for quasi-1D diffusive wires (see, e.g., Ref. \cite{Anthore03}), and constants $c_{1}$ and $c_2$. The length $\ell_B$ is the characteristic length scale for time-reversal symmetry breaking in magnetic field \cite{Altshuler81}, whereas $\ell_M$ and $C_M$ in Eq.~(\ref{l_M}) are associated with the band Berry phase $\beta_M$ controlling the rate of $180^\circ$-backscattering. $\beta_M$ deviates from the universal value of $\pi$, which characterizes gapless Dirac-cones \cite{Ando02}, because of the finite band gap ${\cal M}$ and the quadratic correction to the dispersion ${\cal B}k^2_{_F}>0$. $E_F$ and $k_{_F}=\sqrt{ 2\pi n }$ are the Fermi energy, momentum, and $ {\cal D} k^2_{_F}>0$  is the parabolic part of the QW energy spectrum [see also supplemental online material (SOM)]. $\ell $ in Eq.~(\ref{l_M}) is determined by the carrier mobility and density as $\ell = (h/e) \mu \sqrt{n/(2\pi)}$. It should be noted that Eqs.~(\ref{dS}) and (\ref{l_M}) are valid for weak deviations of $(1 - \beta_M/\pi)^2= ( {\cal M} + {\cal B} k^2_{_F} )^2/(E_F - {\cal D} k^2_{_F}  )^2 \ll 1$.
We use Eq.~(\ref{dS}) to fit the measured magnetic-field and temperature dependence of the WAL for both the normal and the inverted QW. The band structure parameters ${\cal M}$, ${\cal A}$, ${\cal B}$ and ${\cal D}$ of the BHZ model, and the carrier densities, deduced from appropriate Hall measurements, are listed in Tables I and III of the SOM. $c_1$ and $c_2$ are determined by simultaneously fitting the magnetoconductivity data of Fig.~\ref{dS_B} and the temperature dependence of the WAL peaks presented in Fig.~\ref{dS_T}a (see also Table~III of the SOM). The obtained fits are plotted with the experimental data in Figs.\ 2 and 3.
%
%%%   Figure 3   %%%
%
%
%

%%%%%%%%%%%%%%%%%%%%%%%%%%   Discussion   %%%%%%%%%%%%%%%%%%%%%%%%%%%%%%%%%%

{\em Discussion} --
From the quality of the fits it is evident that the quasi-1D diffusive transport model represents the experimental data remarkably well. Additionally, the temperature-dependent dephasing length $\ell_\varphi(T)$ can be extracted, which is shown in \ref{dS_T}b for both inverted and normal QWs. At the lowest experimental temperature, $T=1.8$ K, $\ell_\varphi(T)$ is well over 1200 nm for both samples, which significantly exceeds the width, $w=370$ nm, of the wires and adds to the justification of the conjecture that the observed WAL effect is quasi-1D. It should be noted that comparable dephasing lengths were reported for narrow Ga$_x$In$_{1-x}$As/InP wires \cite{Schaepers06}.
We now turn to the observed difference in the WAL amplitude between the inverted and normal QWs (cf. Fig.~\ref{dS}). The large value of $\ell_\varphi(T)$ for the normal QW rules out that strong dephasing can explain the less pronounced WAL. Moreover, magnetotransport experiments on single wires (not shown here) show universal conductance fluctuations equally pronounced in both types of samples, which is another clear evidence for similar phase-coherent transport properties.

In quasi ballistic wires, the WAL amplitude can also be affected by multiple boundary scattering due to partial magnetic flux cancellation. This occurs when the 1D mean-free path, $\ell_{1D}$, is larger than width $w$. From the zero field conductivities of the wires and Hall bars
we estimate $\ell_{1D}$ to be about 100 nm and 50 nm for the inverted and normal QW structrue, respectively. Since for both samples $\ell_{1D}$ is smaller than the width $w$ of the wires, boundary scattering can hardly account for the experimental observation either. Hence, we conclude that the weaker WAL in wires with a normal band order is related to the specifics of the band structure that are encoded in the Berry phase, $\beta_M$.
The inset in Fig.~3b shows the  Berry phase $\beta_M$ extracted from fitting the experimental data for the inverted and normal QW  to Eqs.~(1)-(3). For both QWs,  $\beta_M$ deviates from $\pi$ because of the non-zero total gap ${\cal M} + {\cal B} k^2_{_F}$ in both wells. Interestingly, the deviation from $\pi$ is stronger for the normal QW. For a normal band structure, such a non-zero, positive deviation is expected because the terms ${\cal M}$ and ${\cal B} k^2_{_F}$ in Eq.~(4) are  both positive and thus can never cancel each other, whereas a partial or complete cancellation of both terms occurs for the inverted QW because of the negative gap parameter, ${\cal M} < 0$. Consequently, normal QWs may show enhanced backscattering and a tendency to WL, which is accounted for by the negative subleading correction $C_M$ in Eq.~(\ref{dS}), reducing the WAL effect.
Formally, the subleading correction $C_M$ in Eqs.~(1) and (3) reflects the axial spin-rotation symmetry of the BHZ model. Depending on the parameters of the BHZ model, the Berry phase $\beta_M$ now has a value between $\pi$ and $2\pi$, which results a reduced WAL compared with the simplectic case where $\beta_M = \pi$ and $\delta\sigma_{xx} = (2e^2/h) (\ell_\varphi/w)$ (see also SOM).

In agreement with our analytical results, a weaker WAL for the normal band gap QW structures has been also found in numerical studies for HgTe QWs within the BHZ model \cite{Krueckl12}. It is worth noting that the spin-rotation symmetry of the BHZ model can be broken by the SOI arising from bulk or/and structural inversion asymmetry \cite{Rothe10}. Strong SOI-induced spin mixing may in general obscure a manifestation of Berry phase effects related to the Dirac band structure \cite{Krueckl12,Ostrovsky12}. However, in the present experiments, the Rashba SOI is adjusted to zero or, at least, to be very weak. As to the bulk-inversion asymmetry, it is most probably also negligible because the magneto resistance data do not reveal any corresponding signatures. This implies that calculations based on the BHZ model should give a valid description of our data, as is also
evidenced by the quality of the fits in Figs. 2 and 3.

In conclusion, the observed WAL effect indicates that quantum transport in quasi-1D HgTe QW structures is governed by a Dirac-like single-valley band dispersion. In this respect, the HgTe nanowires are distinct both from graphene nanoribbons, where the edge roughness generates intervalley scattering, and from low-dimensional semiconductors with SOI-split parabolic bands, e.g. InAs nanowires \cite{Hansen05} and Ga$_x$In$_{1-x}$As/InP nanowires \cite{Schaepers06}. The latter systems show a crossover to the WL either with gate voltage \cite{Hansen05} or in narrow wires with $w \sim 300$ nm \cite{Schaepers06}. In both cases, the crossover is attributed to an increasing SOI-induced spin-relaxation length $\ell_{SO}$. The absence of such a crossover in the measurements presented here points again to the dominant role of the band structure that yields a robust WAL even for $\ell_{SO} \gg \ell_\varphi$, when a conventional system display only WL \cite{Hikami80}.

\acknowledgments

This work was supported by the German Research Foundation DFG (SPP 1285 `Halbleiter Spintronik'
and DFG-JST joint research project `Topological Electronics') and the EU ERC-AG program (Project 3-TOP). The authors thank P.M. Ostrovsky, I.V. Gornyi, A.D. Mirlin, and K.\ Richter for fruitful discussions.


\begin{thebibliography}{99}

\bibitem{Bergmann82}
G. Bergmann, Sol. State Commun. {\bf 42}, 815 (1982).

\bibitem{Dresselhaus92}
P.D. Dresselhaus, C.M.A. Papavassiliou, R.G. Wheeler, and R.N. Sacks, Phys. Rev. Lett. {\bf 68}, 106 (1992).

\bibitem{Ando02}
H. Suzuura and T. Ando, Phys. Rev. Lett. {\bf 89}, 266603 (2002).

\bibitem{GeimKim05}
K. S. Novoselov, A. K. Geim, S. V. Morozov, D. Jiang, M. I. Katsnelson, I. V. Grigorieva, S. V. Dubonos, and A. A. Firsov, Nature {\bf 438}, 197 (2005);
Y. Zhang, J. W. Tan, H. L. Stormer and P. Kim, Nature {\bf 438}, 201 (2005).

\bibitem{Geim06}
S.V. Morozov, K. S. Novoselov, M. I. Katsnelson, F. Schedin, L. A. Ponomarenko, D. Jiang, and A. K. Geim, Phys. Rev. Lett. {\bf 97}, 016801 (2006).

\bibitem{Tikhonenko09}
F.V. Tikhonenko, A.A. Kozikov, A.K. Savchenko, R.V. Gorbachev, Phys. Rev. Lett. {\bf  103}, 226801 (2009).

\bibitem{McCann06}
E. McCann, K. Kechedzhi, V.I. Fal'ko, H. Suzuura, T. Ando, and B.L. Altshuler,
Phys. Rev. Lett. {\bf 97}, 146805 (2006).


\bibitem{Minke12} % G ribbons experiment
S. Minke, J. Bundesmann, D. Weiss, and J. Eroms, Phys. Rev. B {\bf 86}, 155403 (2012).

\bibitem{Ortmann11} % G ribbons theory
F. Ortmann, A. Cresti, G. Montambaux, and S. Roche, Europhys. Lett. {\bf 94}, 47006 (2011).

\bibitem{Bernevig06}
B. A. Bernevig and T. L. Hughes and S. C. Zhang, Science {\bf 314}, 1757 (2006).

\bibitem{Koenig07}
M. K{\"o}nig, S. Wiedmann, C. Br{\"u}ne, A. Roth, H. Buhmann, L. W. Molenkamp, X.-L. Qi and S.-C. Zhang,
Science {\bf 318}, 766 (2007).


\bibitem{Buettner11}
B. B\"uttner, C.X. Liu, G. Tkachov, E.G. Novik, C. Br\"une, H. Buhmann, E. M. Hankiewicz, P. Recher, B. Trauzettel, S. C. Zhang and L. W. Molenkamp,
Nature Phys. {\bf 7}, 418  (2011).

\bibitem{Tkachov11a}
G. Tkachov, C. Thienel, V. Pinneker, B. B\"uttner, C. Br\"une, H. Buhmann, L. W. Molenkamp, and E. M. Hankiewicz,
Phys. Rev. Lett. {\bf 106}, 076802 (2011).

\bibitem{Zholudev12}
J. N. Hancock, J. L. M. van Mechelen, A. B. Kuzmenko, D. van der Marel, C. Br\"{u}ne, E. G. Novik, G. V. Astakhov, H. Buhmann, and L. W. Molenkamp, Phys. Rev. Lett. {\bf 107}, 136803 (2011).

\bibitem{Olbrich13}
A.M. Shuvaev, G. V. Astakhov, G. Tkachov, C. Br\"{u}ne, H. Buhmann, L. W. Molenkamp, and A. Pimenov, Phys. Rev. B {\bf 87},  121104(R) (2013).

\bibitem{Olshan10}
E. B. Olshanetsky, Z. D. Kvon, G. M. Gusev, N. N. Mikhailov, S. A. Dvoretsky, and J. C. Portal, JETP Lett. {\bf 91}, 347 (2010).

\bibitem{Minkov12}
G. M. Minkov, A. V. Germanenko, O. E. Rut, A. A. Sherstobitov, S. A. Dvoretski, and N. N. Mikhailov, Phys. Rev. B {\bf 85}, 235312 (2012).


%\bibitem{Gui04}
%Y.S. Gui, C.R. Becker, N. Dai, J. Liu, C.J. Qui, E.G. Novik, M. Sch\"afer, X.Z. Shu, H.J. Chu, H. Buhmann, L.W. Molenkamp, Phys. Rev. B {\bf  70}, 115328 (2004).

%\bibitem{Novik05}
%E.G. Novik, A. Pfeuffer-Jeschke, T. Jungwirth, V. Latussek, C.R. Becker, G. Landwehr, H. Buhmann, and L.W. Molenkamp, Phys. Rev. B {\bf 72}, 035321 (2005).


\bibitem{Tkachov11b}
G. Tkachov and E.M. Hankiewicz, Phys. Rev. B {\bf 84}, 035444 (2011).

\bibitem{Anthore03}
F. Pierre, A. B. Gougam, A. Anthore, H. Pothier, D. Esteve, and N. O. Birge,
Phys. Rev. B {\bf 68}, 085413 (2003).

\bibitem{Altshuler81}
B. L. Altshuler and A. G. Aronov, Pis'ma Zh. Eksp. Teor. Fiz. {\bf 33}, 515 (1981) [JETP Lett. {\bf 33}, 499 (1981)].

\bibitem{Schaepers06} % GaInAs stripes
Th. Sch\"{a}pers, V. A. Guzenko, M. G. Pala, U. Z\"{u}licke, M. Governale, J. Knobbe, and H. Hardtdegen,
Phys. Rev. B {\bf 74}, 081301(R) (2006).


\bibitem{Krueckl12}
V. Krueckl and K. Richter, Semicond. Sci. Technol. {\bf 27}, 124006 (2012).

\bibitem{Rothe10}
D. G. Rothe, R. W. Reinthaler, C.-X. Liu, L. W. Molenkamp, S.-C. Zhang and E. M. Hankiewicz,  New J. Phys. {\bf 12}, 065012 (2010).

\bibitem{Ostrovsky12}
P. M. Ostrovsky, I. V. Gornyi, and A. D. Mirlin, Phys. Rev. B {\bf 86}, 125323 (2012).


\bibitem{Hansen05} % InAs nanowires
A. E. Hansen, M. T. Bj\"{o}rk, C. Fasth, C. Thelander, and L. Samuelson,
Phys. Rev. B {\bf 71}, 205328 (2005).


\bibitem{Hikami80}
S. Hikami, A. I. Larkin, and N. Nagaosa, Prog. Theor. Phys. {\bf 63}, 707 (1980).

\end{thebibliography}
\end{document}